\documentclass[11pt]{llncs}

 \usepackage{amssymb,amsfonts,amsmath}
\usepackage{xspace}

\newcommand{\HSP}{\textsc{hsp}\xspace}

\def\eqref#1{(\ref{#1})}
\newcommand{\E}{\mathbb{E}}
\newcommand{\F}{\mathbb{F}}
\newcommand{\Z}{\mathbb{Z}}
\newcommand{\C}{\mathbb{C}}

\def\cS{{\mathcal S}}

\def\cX{{\mathcal X}}

\def\fl#1{\left\lfloor#1\right\rfloor}
\def\rf#1{\left\lceil#1\right\rceil}

\begin{document}

\title{Classical and Quantum Algorithms for Exponential Congruences}

\author{Wim van Dam\inst{1} \and Igor E.~Shparlinski\inst{2}}

\institute{Department of Computer Science, Department of Physics,
University of California, Santa Barbara, CA 93106-5110, USA \\
\email{vandam@cs.ucsb.edu} 
\and 
Department of Computing, Macquarie University, NSW 2109, Australia\\
\email{igor@ics.mq.edu.au} 
}

\maketitle

\begin{abstract}
  We discuss classical and quantum algorithms for solvability testing
  and finding integer solutions $x,y$ of equations of the form $af^x +
  bg^y = c$ over finite fields $\F_q$.  A quantum algorithm with time
  complexity $q^{3/8} (\log q)^{O(1)}$ is presented. While still
  superpolynomial in $\log q$, this quantum algorithm is significantly
  faster than the best known classical algorithm, which has time
  complexity $q^{9/8} (\log q)^{O(1)}$. Thus it gives an example of a
  natural problem where quantum algorithms provide about a cubic
  speed-up over classical ones.
\end{abstract}



\section{Introduction}

Let $\F_q$ be a finite field of $q$ elements and let $ \F_q^*$ denote
the multiplicative group of nonzero elements of $\F_q$.  For
$a,b,c,f,g \in \F_q^*$ we consider the equations
\begin{equation}
\label{eq:Equation}
af^x + bg^y = c
\end{equation}
in nonnegative integers $x$ and $y$.

Equation~\eqref{eq:Equation} has a long history of study in number
theory. In particular, it is dual closely related to the classical
problem of finding $f,g\in\F_q$ for fixed $a,b$ and $x,y$ from the
theory \emph{cyclotomic classes}, see~\cite{BeEvWi,Stor}, which looks
like a dual problem to studying Equation~\eqref{eq:Equation} but in
fact, after a change of variables, become equivalent.

Furthermore, Equation~\eqref{eq:Equation} and variants of it also
appeared in recent work of A.~Lenstra and B.~de~Weger~\cite{LeWe} and
have been shown to be of cryptographic significance.  In particular,
the question about difficulty of finding solutions to
Equation~\eqref{eq:Equation} has been discussed in~\cite{LeWe} but now
concrete results have been know before the present work.

In the theory of quantum computing the task of finding the solutions
to Equation~\eqref{eq:Equation} is of importance when trying to solve the
\emph{hidden subgroup problem} for semi-direct product groups $\Z/N
\rtimes \Z/p$ with $p= \Theta(\sqrt{N})$, see~\cite{BCvD},
where, as usual, $A =  \Theta(B)$ means that $A = O(B)$ and
$B = O(A)$ (hereafter all implied constants are absolute).
Furthermore it is also natural to consider this problem as a
generalization of the discrete logarithm problem in $\F_q$, which can
be solved efficiently using Shor's algorithm~\cite{Shor}.

In this article we use some number theoretic tools to design classical
and quantum algorithms that are more efficient than the brute force
search (but unfortunately still have a running time exponential in the
input size $\log q$).  We use our classical algorithm to measure the
level of improvement that can be achieved by allowing quantum
algorithms.  Ignoring $\log q$ terms, the classical algorithm that we
present here has complexity ${O}^*(q^{9/8})$ (which seems to be
the best known) whereas we also present a quantum algorithm with
complexity ${O}^*(q^{3/8})$, where, as usual, $A = {O}^*(B)$
means that $A = B (\log B)^{O(1)}$.  In particular, it gives an
example of a natural problem where quantum algorithms provide an
asymptotically cubic speed-up over classical ones.

Certainly if $f$ or $g$ is a primitive root, which generates all of
$\F_q^*$, then the problem is not harder than the discrete logarithm
problem. Moreover, in general our results suggest that finding
solutions to Equation~\eqref{eq:Equation} becomes easier in case
$f$ or $g$ is of large order, but still it appears to be much harder
than the discrete logarithm problem.

\section{The Number of Solutions to the Equation}

\subsection{The Worst Case}
\label{sec:W-case}

Here we use bounds of multiplicative character sums over finite fields
to show that if the orders of $f$ and $g$ are large enough, then
Equation~\eqref{eq:Equation} has a solution with at least one
reasonably small component $x$ or $y$.

\begin{lemma}    \label{lem:Large Orders}
  Let $a,b,c\in \F_q^*$ and let $f$ and $g \in \F_q$ be of multiplicative
  orders $s$ and $t$, respectively.  Then for any positive integer
  $r\leq t$, the equation $a f^x+bg^y=c$  has $rs/(q-1) +
  O(q^{1/2}\log q)$ solutions in nonnegative integers $x$ and $y$ with
$x\in\{0,\dots,s-1\}$ and $y\in\{0,\dots,r-1\}$.
\end{lemma}

\begin{proof} 
  Let $k = (q-1)/s$ and let $\cX_k$ be the group of all $k$
  multiplicative characters $\chi:\F_q\rightarrow \C$ of order $k$,
  that is, $\chi^k = \chi_0$, the principal character, for any
  $\chi\in \cX_k$ (see~\cite{LN}). Note that for all non-empty $\cX_k$
  this group contains $k$ elements.  For any $u \in \F_q$ we have
\begin{equation*}
\frac{1}{k}\sum_{\chi \in \cX_k} \chi(u) =
\begin{cases}
   1 , &  \quad \mbox{if} \
u^s  = 1,\\
0, &  \quad  \mbox{otherwise.}
\end{cases}
\end{equation*}
Noting that $u \in \F_q$ belongs to the group generated by $f$ if
and only if $u^s =1$, 
we derive that the number $N_{a,b,c}(r,s)$ of solutions
to Equation~\eqref{eq:Equation} with $x\in\{0,\dots,s-1\}$
and $y\in\{0,\dots, r-1\}$ equals
\begin{equation*}
N_{a,b,c}(r,s) = \sum_{y =0}^{r-1} \frac{1}{k}\sum_{\chi \in \cX_k} \chi( a^{-1}(c -bg^y)).
\end{equation*}
Changing the order of summation and separating the term $r/k$ corresponding to
the principal character $\chi_0$ we obtain
\begin{equation*}
\Big|N_{a,b,c}(r,s)- \frac{r}{k}\Big| \leq
\frac{1}{k}\sum_{\chi \in \cX_k\setminus \{\chi_0\}} \chi(a^{-1})\sum_{y =0}^{r-1}
\chi(c-bg^y).
\end{equation*}
By~\cite[Theorem~3]{Yu} (see also~\cite{DobWil}) each summation
over $y$ is bounded by $O(q^{1/2} \log q)$ (provided $1 \leq r \leq t$),
hence we have
\begin{equation*}
N_{a,b,c}(r,s) = \frac{r}{k} + O(q^{1/2}\log q),
\end{equation*}
which concludes the proof.
\qed
\end{proof}

{From} Lemma~\ref{lem:Large Orders} we can immediately conclude the following.

\begin{corollary}   
 \label{cor:Existence}
  Let $a,b,c\in\F_q^*$ and let $f$ and $g \in \F_q$ be of multiplicative orders
  $s$ and $t$, respectively.  There exists an absolute constant $C >
  0$ such that if for some integer $r$ we have
\begin{equation*}
C  q^{3/2} s^{-1} \log q \leq r \leq t,
\end{equation*}
then the equation  $a f^x+bg^y=c$
has  a solution in integers $x$ and $y$ with
$x\in\{0,\dots,s-1\}$ and
$y\in\{0,\dots,r-1\}$.
\end{corollary}
We remark that the constant $C$ in Corollary~\ref{cor:Existence} is
independent of all variables involved ($a,b,c,f,g$ and $q$) and that
it is effectively computable.  This result reduces the number of
$(x,y)$ pairs that has to be searched for a solution to
Equation~\eqref{eq:Equation}.  In Sections~\ref{sec:ClasAlgWorst} and
\ref{sec:QAlgWorst} we show how this reduction can be used to
construct non-trivial worst case algorithms, both classical and quantum.

\subsection{The Typical Case} \label{sec:A-case} 
To solve the equation $af^x + bg^y=c$ for typical $c\in\F_q$ we now
show that for almost all $c \in \F_q^*$ the results of
Corollary~\ref{cor:Existence} can be improved, which in turn will
yield better average case algorithms for the central problem.

\begin{lemma}   
 \label{lem:Large Orders-Av}
 Let $a,b,c\in \F_q^*$ and let $f$ and $g \in \F_q$ be of
 multiplicative orders $s$ and $t$, respectively.  For any positive
 integer $r\leq t$ and $\delta>0$, for all but $q/\delta^2$ elements
 $c \in \F_q^*$, the equation $a f^x+bg^y=c$ has $rs/q+\vartheta
 \delta \sqrt{r}$ solutions in nonnegative integers $x$ and $y$ with
 $x\in\{0,\dots,s-1\}$, $y\in\{0,\dots,r-1\}$ and $-1<\vartheta<
 1$. 
\end{lemma}

\begin{proof} Let $\psi:\F_q\rightarrow \C$ be a nontrivial additive
  character.  We recall that for for any $u \in \F_q$ we have
\begin{equation*}
\frac{1}{q}\sum_{\lambda\in \F_q}\psi(\lambda u) =
\begin{cases}
1,&\quad\mbox{if $v = 0$,}\\
0,&\quad\mbox{if $v\in \F_q^*$.}
\end{cases}
\end{equation*}
As in the proof of Lemma~\ref{lem:Large Orders} we use
$N_{a,b,c}(r,s)$ to denote the number of solutions to
Equation~\eqref{eq:Equation} with $x\in\{0,\dots,s-1\}$ and
$y\in\{0,\dots, r-1\}$. We have 
\begin{align*}
N_{a,b,c}(r,s) & = \sum_{x=0}^{s-1} \sum_{y =0}^{r-1} \frac{1}{q}\sum_{\lambda\in \F_q} 
\psi( \lambda(a f^x+bg^y-c))\\
& = \frac{sr}{q} +  \frac{1}{q}\sum_{\lambda\in \F_q^*}  \sum_{x=0}^{s-1} \sum_{y =0}^{r-1}
\psi( \lambda(a f^x+bg^y-c)),  
\end{align*}
which averaged over $c \in \F_q$ equals $sr/q$. To calculate the variance from its average, 
we look at the value defined by 
\begin{equation*}
W_{a,b}(r,s) = \sum_{c\in \F_q} \Big(N_{a,b,c}(r,s)- \frac{rs}{q}\Big)^2, 
\end{equation*}
which equals 
\begin{multline*}
\frac{1}{q^2}\sum_{c\in \F_q}  \sum_{\lambda_1,\lambda_2\in \F_q^*}  \sum_{x_1,x_2=0}^{s-1} 
\sum_{y_1,y_2 =0}^{r-1} \psi( \lambda_1(a f^{x_1}+bg^{y_1}-c) +  \lambda_2(a f^{x_2}+bg^{y_2}-c))\\
   = \frac{1}{q^2}\sum_{\lambda_1,\lambda_2\in \F_q^*}  \sum_{x_1,x_2=0}^{s-1} \psi( a(\lambda_1
f^{x_1}+\lambda_2 f^{x_2}))
 \sum_{y_1,y_2 =0}^{r-1}\psi( b(\lambda_1g^{y_1} +  \lambda_2g^{y_2}))\times \\
 \qquad \sum_{c\in \F_q}\psi( c(\lambda_2+\lambda_1)). 
\end{multline*} 
The inner sum over $c$ vanishes unless $\lambda_1 =-\lambda_2$ (in
which case it is $q$) and therefore
\begin{align*}
W_{a,b}(r,s)
&  = \frac{1}{q}\sum_{\lambda\in \F_q^*}  \sum_{x_1,x_2=0}^{s-1} \psi( a\lambda(f^{x_1}-  f^{x_2}))   \sum_{y_1,y_2 =0}^{r-1}\psi( b\lambda(g^{y_1} -  g^{y_2})) \\
 &  = \frac{1}{q}\sum_{\lambda\in \F_q^*} \Bigg| \sum_{x=0}^{s-1} \psi( a\lambda f^{x})\Bigg|^2 
\Bigg|  \sum_{y =0}^{r-1}\psi( b\lambda g^{y})  \Bigg|^2 .
\end{align*}
It is well known that 
\begin{equation*}
\Bigg| \sum_{x=0}^{s-1} \psi( a\lambda f^{x})\Bigg|^2 \le q^{1/2},
\end{equation*}
for example, this follows from~\cite[Theorem~8.78]{LN}
taken with $k=1$ and $g^0,g^1,\dots,g^{s-1}$ the impulse response
sequence (it can also be derived from the bound of Gauss sums,
see~\cite[Theorem~5.32]{LN}).  Therefore
\begin{equation*}
W_{a,b}(r,s)
\leq  \sum_{\lambda\in \F_q}   
\Bigg| \sum_{x=0}^{s-1} \psi( a\lambda f^{x})\Bigg|^2 
\end{equation*}
(note that we have added $\lambda=0$ into the last sum). 
We also have the straightforward equality 
\begin{equation*}
\sum_{\lambda\in \F_q}\Bigg|\sum_{y =0}^{r-1}\psi( b\lambda g^{y}) \Bigg|^2 = 
\sum_{\lambda\in \F_q}\Bigg|\sum_{y =0}^{r-1}\psi( \lambda g^{y}) \Bigg|^2 
= qr
\end{equation*}
(essentially, this is Parseval's identity, i.e.\ we used the unitarity 
of the Fourier transformation over $\F_q$ on the characteristic vector 
of the set $\{g^0,\dots,g^{r-1}\}$) and thus
\begin{equation*}
W_{a,b}(r,s) = 
\sum_{c\in \F_q} \Bigg|N_{a,b,c}(r,s)- \frac{rs}{q}\Bigg|^2  \leq qr.
\end{equation*}
Hence, for any $\delta>0$,  the violation
\begin{equation*}
 \Bigg|N_{a,b,c}(r,s)- \frac{rs}{q}\Bigg| \geq  \delta\sqrt{r}
 \end{equation*}
 holds for no more than $q/\delta^2$ values of $c \in \F_q^*$. \qed 
\end{proof}

Using $\delta=\sqrt{\log q}$ in
Lemma~\ref{lem:Large Orders-Av}, we see that  for all but
$q/\!\log q = o(q)$ elements $c\in\F_q^*$ the equation $a f^x+bg^y=c$
has $rs/q+\vartheta \sqrt{r\log q}$ solutions in  $x\in\{0,\dots,s-1\}$,
$y\in\{0,\dots,r-1\}$ with $-1<\vartheta< 1$.
Therefore we can immediately conclude the
following.

\begin{corollary}      \label{cor:Existence-Av}
  Let $a,b,c\in\F_q^*$ and let $f$ and $g \in \F_q$ be of multiplicative orders
  $s$ and $t$, respectively.  
If for some integer $r$ we have
\begin{equation*}
 q^{2} s^{-2} \log q \leq r \leq t,
\end{equation*}
then for all but $o(q)$ elements $c \in \F_q^*$, the equation $a f^x+bg^y=c$
has  a solution in integers $x$ and $y$ with
$x\in\{0,\dots,s-1\}$ and
$y\in\{0,\dots,r-1\}$.
\end{corollary}  

\section{Classical Algorithms}

\subsection{Worst Case Classical Algorithm} \label{sec:ClasAlgWorst}

We start with a classical deterministic algorithm that
is more efficient than brute search.

\begin{theorem}     \label{thm:Classic}
  Let $a,b,c,f,g\in \F^*_q$.  One can either find a solution
  $x,y\in\Z_{\geq 0}$ of the equation $af^x+bg^y=c$ or decide that it
  does not have a solution in deterministic time 
  $q^{9/8} (\log q)^{O(1)}$ on a classical computer.
\end{theorem}

\begin{proof} 
  Using a standard deterministic factorization algorithm, we factor
  $q-1$ and find the orders $s$ and $t$ of $f$ and $g$ in time
  $q^{1/2} (\log q)^{O(1)}$.  Assume without loss of generality that
  $s\geq t$ (otherwise of the roles of $s$ and $t$ are reversed in the
  proof below).  Let $C$ be the constant of
  Corollary~\ref{cor:Existence} and define
\begin{equation}\label{eq:defr}
r = \rf{C  q^{3/2} s^{-1} \log q }.
\end{equation}
By Corollary~\ref{cor:Existence}, if $r\leq t$
then the central equation $af^x+bg^y=c$ is solvable for 
$(x,y) \in \{0,\dots,s-1\} \times \{0,\dots,r-1\}$. 
Otherwise, if $r>t$, there may or may not be a solution 
with $(x,y) \in \{0,\dots,s-1\} \times \{0,\dots,t-1\}$.  
As a result, the following algorithm proves the theorem.

If $r \leq t$ then for every $y\in\{0,\dots,r-1\}$ we evaluate
$a^{-1}(c - bg^x )$ and then try to compute its discrete logarithm to
base $f$, that is, an integer $x$ with $f^x = a^{-1}(c- bg^y)$, in
deterministic time $s^{1/2} (\log q)^{O(1)}$,
see~\cite[Section~5.3]{CrPom}. When found, the algorithm outputs
$(x,y)$ and terminates.  The condition $t \geq r$ and assumption
$s\geq t$ implies for $s$:
\begin{equation*}
s^2 \geq st \geq sr \geq C  q^{3/2}   \log q, 
\end{equation*}
which gives for the time complexity of this case 
\begin{equation*}
r\cdot s^{1/2}(\log q)^{O(1)} 
= q^{3/2} s^{-1/2} (\log q)^{O(1)} 
 \leq  q^{9/8} (\log q)^{O(1)}. 
\end{equation*} 

If $r > t$ we perform the same procedure for every $y\in\{0,\dots,
t-1\}$.  If none of the $y$ yield a solution, the algorithm reports
that the central equation 
has no solution.  In this case, the condition $t < r$ implies that
\begin{equation*}
st < sr \leq  C  q^{3/2}   \log q
\end{equation*}
and since $t \leq s$, the time complexity of this case is also bounded by 
\begin{equation*}
t\cdot s^{1/2} (\log q)^{O(1)}   \leq   (st)^{3/4}  (\log
q)^{O(1)}  \leq  q^{9/8} (\log
q)^{O(1)}, 
\end{equation*} 
which completes the proof. \qed
\end{proof}

It is natural to ask whether one can design a faster probabilistic
algorithm. For some fields this is indeed possible due to the
existence of subexponential algorithms for computing discrete
logarithms, see~\cite[Section~6.4]{CrPom}.  However in general
probabilistic algorithms do not seem to give any significant advantage
for our problem.

\subsection{Typical Case Classical Algorithm} 

Similarly, using Corollary~\ref{cor:Existence-Av} instead of
Corollary~\ref{cor:Existence} and repeating the arguments of the proof
of Theorem~\ref{thm:Classic} with
\begin{equation}
\label{eq:defr-Av}
r = \rf{q^{2} s^{-2} \log q}
\end{equation}
we obtain that for almost all $c$ a stronger result than 
Theorem~\ref{thm:Classic} holds.
\begin{theorem} \label{thm:Classic-Av} 
  Let $a,b,c,f,g\in \F^*_q$.  For all but $o(q)$ elements $c\in\F_q^*$,
  one can either find a solution $x,y\in\Z_{\geq 0}$ of the equation
  $af^x+bg^y=c$ or decide that it does not have a solution in
  deterministic time $q(\log q)^{O(1)}$ on a classical computer.
\end{theorem}

\section{Quantum Algorithms}

\subsection{Worst Case Quantum Algorithms} \label{sec:QAlgWorst}

On a quantum computer one has the advantage that calculating
discrete logarithms can be done efficiently in time $(\log q)^{O(1)}$.
In combination with the quadratic speed-up of quantum searching
this gives the following quantum algorithm for the central problem.
We start with an algorithm that works for \emph{any} $f$ and $g$.

\begin{theorem}    \label{thm:Quantum-1}
  Let $a,b,c,f,g\in \F^*_q$.  One can either find a solution
  $x,y\in\Z_{\geq 0}$ of the equation $af^x+bg^y=c$ or decide that it
  does not have a solution in  time $q^{3/8} (\log
  q)^{O(1)}$ on a quantum computer.
\end{theorem}

\begin{proof}
  We use Shor's algorithm~\cite{Shor} to compute $s$ and $t$ in
  polynomial time.  Without loss of generality we assume that $s \geq
  t$ and we define $r$ by Equation~\eqref{eq:defr} as in the proof of
  Theorem~\ref{thm:Classic}.  A polynomial time quantum subroutine
  $\cS(y)$ is constructed that, using Shor's discrete logarithm
  algorithm~\cite{Shor}, 
  for a given $y$ either finds and returns the integer $x$ with $f^x =
  a^{-1}(c- bg^x)$ or reports that no such $x$ exists.

If $r \leq t$, then, using Grover's search algorithm~\cite{Grov}, 
we search the subroutines $\cS(y)$ for all
$y\in\{0,\dots,r-1\}$ in time 
\begin{equation*}
 r^{1/2} (\log q)^{O(1)} =  q^{3/4} s^{-1/2} (\log
q)^{O(1)} \leq q^{3/8} (\log
q)^{O(1)}.
\end{equation*}

If $r > t$,  we search the  $\cS(y)$ for all
$y\in\{0,\dots,t-1\}$  in time 
\begin{equation*}
t^{1/2} (\log q)^{O(1)} \leq (st)^{1/4} (\log q)^{O(1)}  
\leq q^{3/8} (\log q)^{O(1)}.
\end{equation*}
As in the proof of Theorem~\ref{thm:Classic}, we conclude that due to
our choice of $r$ we either find a solution to
Equation~\eqref{eq:Equation} or conclude that there is no solution.
\qed
\end{proof}

We now show that if $f$ and $g$ are of large order then there is a
more efficient quantum algorithm.

\begin{theorem}    \label{thm:Quantum-2}
  Let $a,b,c,f,g\in \F^*_q$ and let $f$ and $g$ be of multiplicative
  orders $s$ and $t$, respectively.  There is an absolute constant $C$
  such that if
\begin{equation*}
st > C q^{3/2} (\log q)^{1/2}
\end{equation*}
then one can either find a solution $x,y\in\Z_{\geq 0}$ of the
equation $af^x+bg^y=c$ or decide that it does not have a solution
in time $q^{1/2} (st)^{-1/4} (\log q)^{O(1)}$ on a quantum computer.
\end{theorem}

\begin{proof}
  Assume without loss of generality that $s\geq t$.  It follows from
  the condition of the theorem and Lemma~\ref{lem:Large Orders} that
  for some appropriate constant $C$ and
\begin{equation*}
r = \fl{C q^{3/2} s^{-1} (\log q)^{1/2}} \leq t
\end{equation*}
there are  
\begin{equation*}
\frac{rs}{q-1}+O(q^{1/2}\log q) \geq \frac{rs}{2q}
\end{equation*} 
solutions to
Equation~\eqref{eq:Equation} with $x\in\{0,\dots,s-1\}$ and
$y\in\{0,\dots,r-1\}$.

We now use the version of Grover's search algorithm as described
in~\cite{BBHT} that finds one out of $m$ matching items in a set of
size $r$ using only $O(\sqrt{r/m})$
queries. 
Here we search the subroutines $\cS(y)$ for all $y\in\{0,\dots,r-1\}$
with the promise (which follows from Lemma~ that there are $m =
rs/(q-1) + O(q^{1/2}\log q)$ solutions $(x,y)$.  Because for each
value $y$ there can be at most one value $x\in\{0,\dots,s-1\}$ such
that $af^x+bg^y=c$ there are $m$ different values $y$ for which $\cS$
will report a solution $x$, hence a solution will be found in time
\begin{equation*}
(r/m)^{1/2} (\log q)^{O(1)} =
q^{1/2}s^{-1/2}(\log q)^{O(1)}.
\end{equation*}
Since $s \geq (st)^{1/2}$, this concludes the proof.
\qed
\end{proof}
In particular, the running time of the algorithm of Theorem~\ref{thm:Quantum-2}
is upper bounded by $O(q^{1/8} (\log q)^{O(1)} )$.

\subsection{Typical Case Quantum Algorithms} 

Similarly to the classical case, for almost all $c \in \F_q$ stronger
results than those of Theorems~\ref{thm:Quantum-1} and
\ref{thm:Quantum-2} are possible.  For example, defining again $r$ by
Equation~\eqref{eq:defr-Av} and arguing as in the proof of
Theorem~\ref{thm:Quantum-1}, we obtain the following result.

\begin{theorem}    \label{thm:Quantum-1-Av}
  Let $a,b,c,f,g\in \F^*_q$.  For all but $o(q)$ elements $c \in \F_q^*$, 
  one can either find a solution $x,y\in\Z_{\geq 0}$ of the equation
  $af^x+bg^y=c$ or decide that it does not have a solution in time
  $q^{1/3} (\log q)^{O(1)}$ on a quantum computer.
\end{theorem}

Finally,  taking
\begin{equation*}
r = \fl{q^{2} s^{-2} \log q}
\end{equation*}
 and using   Lemma~\ref{lem:Large Orders}  in the 
 argument  of the proof of Theorem~\ref{thm:Quantum-2}, we 
 see that for almost all $c \in \F_q$ the complexity estimate 
 of Theorem~\ref{thm:Quantum-2} holds for a wider range of $s$ and $t$.   
 
 \begin{theorem}    \label{thm:Quantum-2-Av}
  Let $a,b,c,f,g\in \F^*_q$ and let $f$ and $g$ be of multiplicative
  orders $s$ and $t$, respectively.  
For all but $o(q)$ elements $c \in \F_q^*$, 
if
\begin{equation*}
st >   q^{4/3}( \log q)^{2/3}
\end{equation*}
then one can either find a solution $x,y\in\Z_{\geq 0}$ of the
equation $af^x+bg^y=c$ or decide that it does not have a solution
in time $q^{1/2} (st)^{-1/4} (\log q)^{O(1)}$  on a quantum computer.
\end{theorem}

\section{Connection with the Hidden Subgroup Problem}

The \emph{pretty good measurement} approach~\cite{BCvD} to the Hidden
Subgroup Problem (\HSP) over the non-abelian group $\Z/q\rtimes \Z/p$
with $q$ a prime and $q/p^2= (\log q)^{O(1)}$ shows that the \HSP can
be solved efficiently on a quantum computer if one can efficiently
solve the equation $a f^x+bf^y=c$, where $f$ has multiplicative order
$p$ in $\Z/q$. All algorithms presented in this article have
superpolynomial complexity in $\log q$ and hence fall short of this
goal.

For this restricted problem with $f=g$ and $f$ of order $p \approx
\sqrt{q}$, there are $p^2$ possible solutions $(x,y)$, hence even a
classical algorithm has $O^*(q)$ time complexity instead of
the $O^*(q^{9/8})$ of Theorem~\ref{thm:Classic}. Quantum
mechanically, one can `Grover search' the set of solutions
$x\in\{0,\dots,p-1\}$ in time $O^*(q^{1/4})$, which,
although better than the $O^*(q^{3/8})$ of
Theorem~\ref{thm:Quantum-1}, is still far from polynomial in $\log q$.

\section{Remarks and Open Problems}

We remark that in some finite fields classical subexponential
probabilistic algorithms are possible for the discrete logarithm
problem, see~\cite[Section~6.4]{CrPom}.  In such fields, a version of
Theorem~\ref{thm:Classic} can be obtained with an algorithm that runs
in probabilistic time $q^{3/4 + o(1)}$, which is still much slower that
the quantum algorithm of Theorems~\ref{thm:Quantum-1}
and~\ref{thm:Quantum-2}. We note that although over the last 
several years fast heuristic algorithms
for the discrete logarithm problem have been designed to work over 
any finite field, rigorous subexponential algorithms are know only 
for fields of special types (such as prime fields $\F_p$ or binary 
fields $\F_{2^m}$), see~\cite[Section~6.4]{CrPom} for 
more details. Clearly using probabilistic algorithms
one can also get additional speed up in the classical case if 
the multiplicative orders $s$ and $t$ are large (similar to
Theorems~\ref{thm:Quantum-2}
and~\ref{thm:Quantum-2-Av}).

To try to strengthen the presented results one can consider the analogue 
to Equation~\eqref{eq:Equation} for elliptic curves $\E$ over $\F_q$.  
For example, given two $\F_q$-rational points $F, G \in \E(\F_q)$
and the values $a,b,c\in\F_q$ one can ask for solutions to the equation
\begin{equation*}
a\cdot x([u]F) + b\cdot x([v]G) = c
\end{equation*}
in integers $u$ and $v$ (where $x(Q)$ denotes the $x$-coordinate of a
point $Q\in \E(\F_q)$ in a fixed affine model of $\E$ and $[n]Q$
denotes the $n$-fold sum $Q\oplus Q\oplus \cdots\oplus Q$ in the group
of $\E$).  Using bounds of character sums over subgroups of elliptic
curves, see~\cite{KohShp}, one can obtain full analogues of our
results (in fact at the cost of only typographical changes).  This
case is interesting since in the classical scenario even heuristic
subexponential algorithms are not known.

But above of this all, it still remains an open problem whether or not
there exist efficient quantum algorithms that run in time 
$(\log q)^{O(1)}$ for the determining the integer solutions $x,y$ to the
equation $af^x+bg^y=c$ and even the more restricted version $af^x+bf^y=c$
over $\F_q$.

\subsubsection*{Acknowledgments.}

The authors are grateful to Michele Mosca for useful and encouraging
discussions.

This work was initiated during a very pleasant visit by I.S.\ to the
University of California at Santa Barbara whose hospitality is
gratefully acknowledged.  W.v.D.\ is supported by the Disruptive
Technology Office (\textsc{dto}) under Army Research Office
(\textsc{aro}) contract number
\textsc{w}\oldstylenums{911}\textsc{nf-}\oldstylenums{04}\textsc{-r-}\oldstylenums{0009}
and the \textsc{nsf} \textsc{career} award no.~\oldstylenums{0803963};
I.S. is supported by \textsc{arc} grant
\textsc{dp}\oldstylenums{0556431}.

\end{document}